\documentclass[aps,prl,twocolumn,noshowpacs,superscriptaddress,preprintnumbers,amsmath,amssymb]{revtex4-1}

\usepackage{graphicx}
\usepackage{dcolumn}
\usepackage{longtable}
\usepackage{color}
\usepackage{bm}

\begin{document}


\title{Inter-planar coupling dependent magnetoresistivity in high purity layered metals}



\author{N.\ Kikugawa}
\affiliation{National Institute for Materials Science, Tsukuba, Ibaraki 305-0047, Japan}
\affiliation{National High Magnetic Field Laboratory, Florida
State University, Tallahassee-FL 32310, USA}
\author{P.\ Goswami}
\affiliation{National High Magnetic Field Laboratory, Florida
State University, Tallahassee-FL 32310, USA}
\affiliation{Condensed Matter Theory Center, University of Maryland, College Park, Maryland 20742-4111, USA}
\author{A.\ Kiswandhi}\altaffiliation[Present address: ]
{Department of Physics, University of Texas at Dallas, Richardson 75080, USA}
\affiliation{National High Magnetic Field Laboratory, Florida
State University, Tallahassee-FL 32310, USA}
\author{E.\ S. Choi}
\affiliation{National High Magnetic Field Laboratory, Florida
State University, Tallahassee-FL 32310, USA}
\author{D.\ Graf}
\affiliation{National High Magnetic Field Laboratory, Florida
State University, Tallahassee-FL 32310, USA}
\author{R.\ E.\ Baumbach}
\affiliation{National High Magnetic Field Laboratory, Florida
State University, Tallahassee-FL 32310, USA}
\author{J.\ S.\ Brooks}
\affiliation{National High Magnetic Field Laboratory, Florida
State University, Tallahassee-FL 32310, USA}
\author{K.\ Sugii}\altaffiliation[Present address: ]
{The Institute for Solid State Physics, The University of Tokyo, Kashiwa, Chiba 277-8581, Japan}
\affiliation{National Institute for Materials Science, Tsukuba, Ibaraki 305-0003, Japan}
\author{Y.\ Iida}
\affiliation{National Institute for Materials Science, Tsukuba, Ibaraki 305-0003, Japan}
\author{M.\ Nishio}
\affiliation{National Institute for Materials Science, Tsukuba, Ibaraki 305-0047, Japan}
\author{S.\ Uji}
\affiliation{National Institute for Materials Science, Tsukuba, Ibaraki 305-0003, Japan}
\author{T.\ Terashima}
\affiliation{National Institute for Materials Science, Tsukuba, Ibaraki 305-0003, Japan}
\author{P. M. C. Rourke}
\affiliation{H. H. Wills Physics Laboratory, University of Bristol, Tyndall Avenue, BS8 1TL, United Kingdom}
\author{N.\ E. Hussey}
\affiliation{High Field Magnet Laboratory (HFML-EMFL), Radboud University, Toernooiveld 7, Nijmegen, Netherlands},
\affiliation{Radboud University, Institute of Molecules and Materials, Heyendaalseweg 135, 6525 AJ Nijmegen, the Netherlands}
\author{H.\ Takatsu}\altaffiliation[Present address: ]{Department of Engineering, Kyoto University, Kyoto 615-8510, Japan}
\affiliation{Department of Physics, Tokyo Metropolitan University, Tokyo 192-0397, Japan}
\affiliation{Department of Physics, Graduate School of Science, Kyoto University, Kyoto 606-8502, Japan}
\author{S. Yonezawa}
\affiliation{Department of Physics, Graduate School of Science, Kyoto University, Kyoto 606-8502, Japan}
\author{Y. Maeno}
\affiliation{Department of Physics, Graduate School of Science, Kyoto University, Kyoto 606-8502, Japan}
\author{L.\ Balicas}\email{balicas@magnet.fsu.edu}
\affiliation{National High Magnetic Field Laboratory, Florida
State University, Tallahassee-FL 32310, USA}


\date{\today}

\begin{abstract}
\textbf{The magnetic ﬁeld-induced changes in the conductivity of metals are the subject of intense interest, both for revealing new phenomena and as a valuable tool for
determining their Fermi surface. Here, we report a hitherto unobserved magnetoresistive effect in ultra-clean layered metals, namely a negative longitudinal magnetoresistance that is capable of overcoming
their very pronounced orbital one. This effect is correlated with the inter-layer coupling disappearing for fields applied along the so-called Yamaji angles where the inter-layer
coupling vanishes. Therefore, it is intrinsically associated with the Fermi points in the field-induced quasi-one-dimensional electronic dispersion, implying that it results from
the axial anomaly among these Fermi points. In its original formulation, the anomaly is predicted to violate separate number conservation laws for left- and right-handed chiral-
(e.g. Weyl) fermions. Its observation in PdCoO$_2$, PtCoO$_2$ and Sr$_2$RuO$_4$ suggests that the anomaly affects the transport of clean conductors,
particularly near the quantum limit.}
\end{abstract}
\maketitle


The magneto-conductivity or -resistivity of metals under a uniform magnetic field $H$ is highly dependent on the precise shape of their Fermi surface (FS) and on the orientation of 
the current flow relative to the external field \cite{pippard, pippard2}. This is particularly true for high purity metals at low temperatures whose carriers may execute many 
cyclotronic orbits in between scattering events. However, the description of the magnetoconductivity of real systems in terms of the Boltzmann equation including the Lorentz force, 
the electronic dispersion, and realistic scattering potentials is an incredibly daunting task, whose approximate solutions can only be obtained through over simplifications.  
Despite the inherent difficulty in describing the magnetoresistivity of metallic or semi-metallic systems, it continues to be a subject of intense interest. Indeed, in recent years, 
a number of new magnetoresistance phenomena have been uncovered. For example, although semi-classical transport theory predicts a magnetoresistivity $\rho(H) \propto H^2$, certain 
compounds such as $\beta$-Ag$_2$Te display a linear, non-saturating magnetoresistivity \cite{rosenbaum} which is ascribed to the quantum magnetoresistive scenario \cite{abrikosov}, associated with linearly dispersing Dirac-like bands \cite{silvertelluride}. However, in semi-metals characterized by a bulk Dirac dispersion and extremely high electron mobilities such as Cd$_3$As$_2$, the linear magnetoresistivity develops a weak $H^2$ term as the quality of the sample increases \cite{ong}. Its enormous magnetoresistivity is claimed
to result from the suppression of a certain protection against backscattering channels \cite{ong}. The semi-metal WTe$_2$ was also found to display a very large and non-saturating 
magnetoresistivity which is $\propto H^2$ under fields up to 60 T. This behavior was ascribed to a nearly perfect compensation between the densities of electrons and holes \cite{ali}.
Recently, a series of compounds were proposed to be candidate Weyl semi-metals characterized by
a linear touching between the valence and the conduction bands at several points (Weyl points) of their Brillouin zone \cite{weyl_semimetals}. These Weyl points are predicted
to lead to a pronounced negative magnetoresistivity for electric fields aligned along a magnetic field due to the so-called Adler-Abel-Jackiw axial anomaly \cite{bell,nielsen}.
\begin{figure}[htbp]
\begin{center}
\includegraphics[width = 8.6 cm]{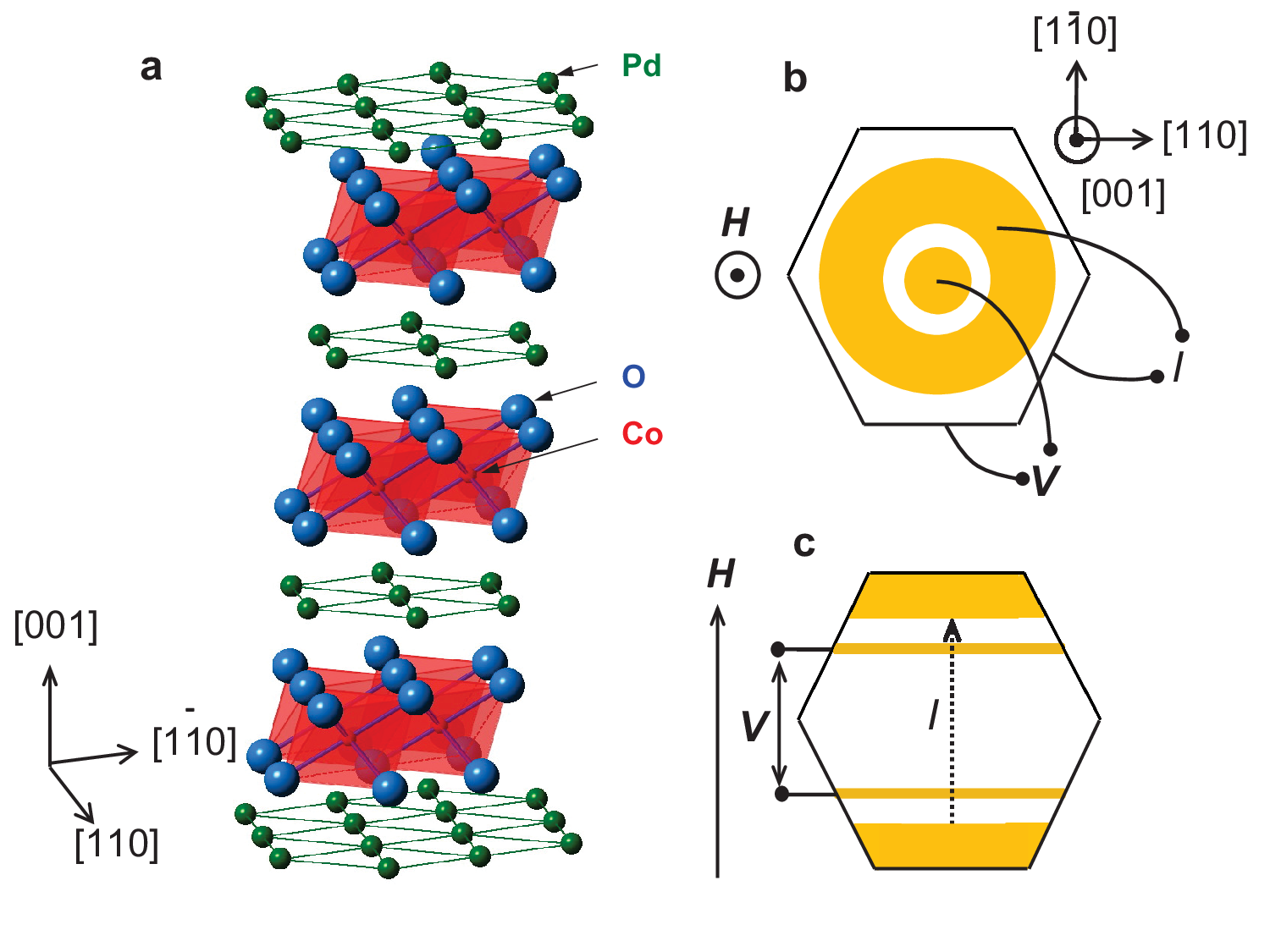}
\caption{\textbf{Crystal structure of PdCoO$_2$ and configuration of electrical contacts.} (\textbf{a}) Crystallographic structure of the langasite
PdCoO$_2$ with Pd, Co and O atoms shown in green, blue, and red, respectively. (\textbf{b}) Configuration of contacts
for measuring the inter-planar longitudinal resistivity ($\rho_{\text{c}}$) showing concentric contacts at the top and at
the bottom surface of each hexagonal platelet like crystal. (\textbf{c}) Configuration of contacts for measuring the
in-plane longitudinal resistivity $(\rho_{[1\overline{1}0]})$ for currents flowing along the $[1\overline{1}0]$-axis and fields applied along
the same direction.}
\end{center}
\end{figure}
\begin{figure*}[htbp]
\begin{center}
\includegraphics[width = 17 cm]{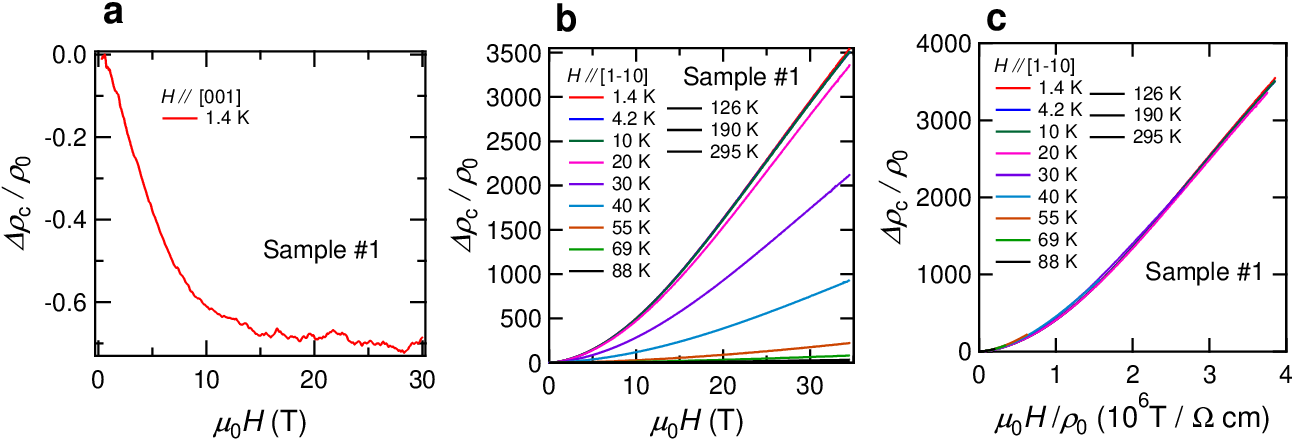}
\caption{\textbf{Negative longitudinal and colossal orbital magnetoresistance of PdCoO$_2$}.
(\textbf{a}) Normalized inter-planar magnetoresistivity $\Delta \rho_{\text{c}}/\rho_0 = (\rho_{\text{c}}(H)- \rho_0)/\rho_0$, where $\rho_0$ is the resistivity at zero-field,
for a PdCoO$_2$ single-crystal and as a function of $H\| j\|c$-axis at $T=1.4$ K. Notice the very pronounced negative longitudinal magnetoresistance arising in the
presence of cyclotron motion in the $ab$ plane. (\textbf{b}) $\Delta \rho_{\text{c}} (H)/\rho_0$  as a function of $H$ applied along the [1$\overline{1}$0] direction and for several temperatures $T$, describing positive transverse magnetoresistance. At $T=1.4$ K, $\Delta \rho_{\text{c}}$ surpasses 350000 \% under a field $H = 35$ T. (\textbf{c}) Kohler scaling of the transverse positive magnetoresistance $\Delta \rho_{\text{c}} (H)$. Notice that i) all data collapse on a single curve as a function of $\mu_0H/\rho_0$ and ii) at low fields $\Delta \rho_{\text{c}} (H)/ \rho_0 \propto (H/\rho_0)^2$ as expected for classical orbital magnetoresistance.}
\end{center}
\end{figure*}

Here, we unveil the observation of yet another magnetoresistive effect, namely a pronounced negative magnetoresistivity in extremely clean and non-magnetic layered metals. 
We study the langasites PtCoO$_2$ and PdCoO$_2$ which are characterized by a single FS sheet and, as with Cd$_3$As$_2$, can display residual resistivities on the 
order of a just few tenths of n$\Omega$cm. Given its extreme low level of disorder, for specific field orientations along which the inter-layer coupling vanishes, PdCoO$_2$ can 
display a very pronounced positive magnetoresistivity which exceeds 550000\% for $H \simeq 35$ T and for currents along the inter-layer axis. Nevertheless, as soon as the field is 
rotated away from these specific orientations and as the field increases, this large orbital effect is overwhelmed  by the emergence of a pronounced negative magnetoresistivity. 
For fields along the inter-layer direction, a strong longitudinal negative magnetoresistivity (LNMR) is observed from $H=0$ T to fields all the way up to $H = 35$ T. 
Very similar behavior is observed in the PtCoO$_2$ compound. For the correlated Sr$_2$RuO$_4$ the LNMR-effect is also observable but only in the cleanest samples, i.e., those 
displaying the highest superconducting transition temperatures. We suggest that this new effect might result from the axial anomaly between Fermi points in a field-induced, 
quasi-one-dimensional electronic dispersion.

\section{Results}
As shown in the Fig. 1a,  PdCoO$_2$ crystallizes in the space group $R\overline{3}m(D^5_{\text{3d}})$, which results from the stacking
of monatomic triangular layers~\cite{takatsu1}. The synthesis of PdCoO$_2$ single crystals is described in the Methods section.
According to band structure calculations~\cite{band1,band2,band3}, the Fermi level $E_{\text{F}}$ is placed between the filled $t_{\text{2g}}$ and the empty $e_{\text{g}}$ levels
with the Pd triangular planes dominating the conductivity, and leading to its
highly anisotropic transport properties. The reported room temperature in-plane resistivity $\rho_{\text{ab}}$
is just $2.6$ $\mu\Omega$cm, making PdCoO$_2$ perhaps the most conductive oxide known to date \cite{hicks}. Figs. 1b and 1c
show the configuration of contacts used for measuring the longitudinal magnetoresistivity of all compounds.
de Haas van Alphen (dHvA) measurements~\cite{hicks} reveal a single, corrugated and nearly two-dimensional Fermi
surface (FS) with a rounded hexagonal cross-section, in broad agreement with both band structure calculations \cite{band1,band2,band3}
and angle resolved photoemission measurements~\cite{arpes}. dHvA yields an average Fermi wave-vector $\overline{k_\text{F}} = \sqrt{2e\overline{F}/\hbar} =  9.5  \times 10^9$ m$^{-1}$ or an average Fermi velocity $\overline{v_\text{F}}= \hbar \overline{k_\text{F}} / \mu =7.6 \times 10^5$ m/s
(where $\mu \simeq 1.5$ is the carrier effective mass~\cite{hicks} in units of free electron mass).
Recent measurements of inter-planar magnetoresistivity $\rho_{\text{c}} (H)$
reveal an enormous enhancement for fields along the [1$\overline{1}$0]-direction, i.e., increasing by $\sim 35 000$ \% at 2 K under $H = 14$ T, which does not follow the characteristic $\rho(H) \propto H^2$ dependence at higher fields~\cite{takatsu2}. This behavior can be reproduced
qualitatively by semi-classical calculations, assuming a very small scattering rate~\cite{takatsu2}.
Most single-crystals display in-plane residual resistivities $\rho_{\text{ab0}}$ ranging from only
$\sim 10$ up to $\sim 40$ n$\Omega$cm, which correspond to transport lifetimes $\tau_{\text{tr}} = \mu/ne^2\rho_{\text{ab0}}$ ranging from $\gtrsim 20$ down
to $ \simeq 5.5$ ps ($e$ is the electron charge and $n\simeq 2.4 \times 10^{28}$ m$^{-3}$ \cite{takatsu1}) or mean free paths $\ell = v_{\text{F}} \tau_{\text{tr}}$
ranging from $\sim 4$ up to 20 $\mu$m~\cite{hicks}. However, according to Ref.~\onlinecite{hicks} the quasiparticle  lifetime  $\tau$ extracted from
the Dingle temperature becomes (in units of length) $\overline{v_\text{F}} \tau \sim 0.6$ $\mu$m. Hence, the transport lifetime is larger than the quasiparticle lifetime by at
least one order of magnitude, which is the hallmark of a predominant forward scattering mechanism, see Ref. \onlinecite{Goswami}. For a magnetic field along
$c$ axis, $\omega_{\text{c}} \tau_{\text{tr}}>1$ when $H \gtrsim 1$ T, in contrast $\omega_{\text{c}} \tau>1$ when $H>10$ T. These estimations suggest the importance of the Landau quantization for
understanding our observations over a wide range of fields up to $H \sim 30$ T.

As shown in Figure 2a, the low-$T$ magnetoresistivity or $\Delta \rho_{\text{c}} = (\rho_{\text{c}}-\rho_0)/\rho_0$, where $\rho_0$ is the zero-field inter-planar resistivity, decreases (up to $\sim 70$ \%) in a magnetic field of 30 T oriented parallel to the applied current. Given that PdCoO$_2$ is non-magnetic and extremely clean (see Methods), this effect cannot be attributed to magnetic impurities. In addition, given the magnitude of the effect it cannot be explained in terms of the suppression of weak localization~\cite{antilocalization,Bergmann}. To support both statements, in Fig. 2b we show $\Delta \rho_{\text{c}}$ for a PdCoO$_2$ single-crystal as a function of $H$
applied  along the $[1\overline{1}0]$ planar-direction and for several temperatures $T$. In sharp contrast to results shown in Fig. 2a, as $T$ decreases $\Delta \rho_{\text{c}} (H)$ increases considerably, by more than 3 orders of magnitude when $T < 10$ K, thus confirming the absence of scattering by magnetic impurities or any role for weak localization. Notice also that $\Delta \rho_{\text{c}} \propto H^2$ at low fields, which indicates that the inter-layer transport is coherent at low fields\cite{ross}. Figure 2c depicts a simple Kohler plot of the magnetoresistivity shown in b, where the field has also been
normalized by $\rho_0(T)$, which indicates unambiguously, that the transverse magnetoresistive effect in PdCoO$_2$ is exclusively orbital in character and is dominated by the
scattering from impurities/imperfections and phonons \cite{pippard}.
\begin{figure*}[htb]
\begin{center}
\includegraphics[width = 18 cm]{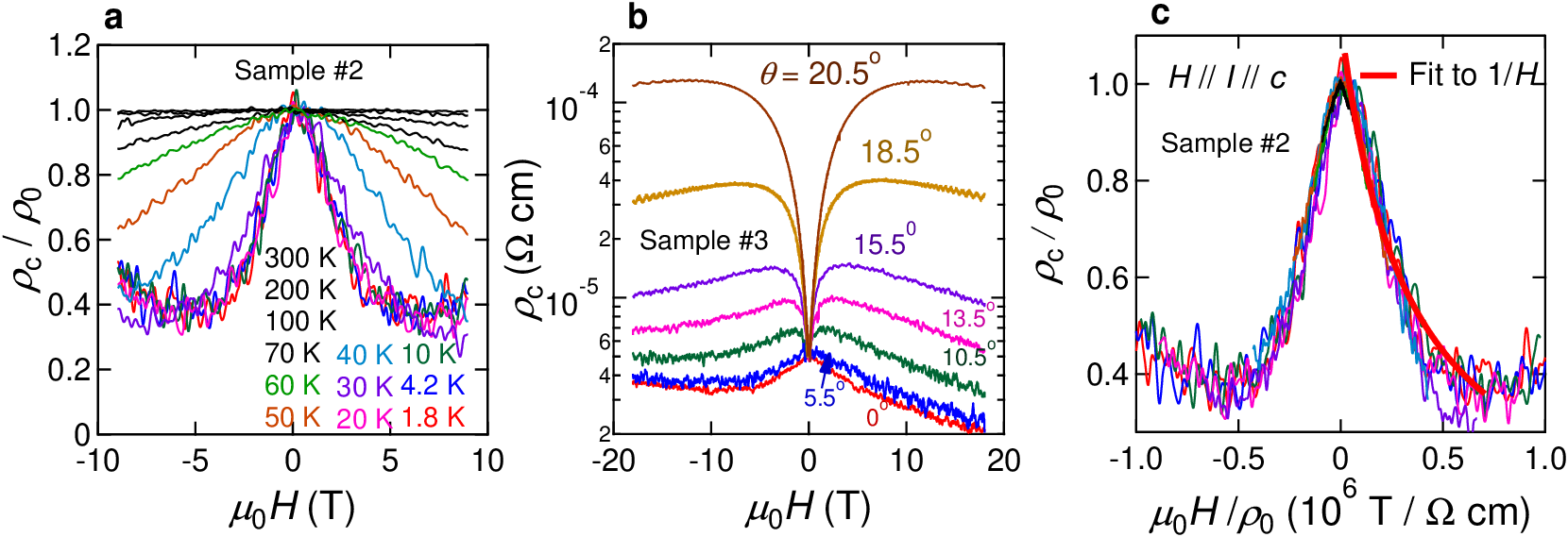}
\caption{\textbf{Anomalous magnetoresistive response of PdCoO$_2$.}
(\textbf{a}) inter-layer resistivity $\rho_{\text{c}}$ normalized by its zero-field value $\rho_0$ as a function of the external field $\mu_0H$ and for $\mu_0H$ parallel to current
$I$ (itself parallel to the sample inter-layer \emph{c}-axis) and for several temperatures $T$. Notice the very pronounced negative magnetoresistivity, that is $\rho_{\text{c}}/\rho_0$
decreases by a factor $> 60$\% when sweeping the field from 0 to 5 T. Notice also that this effect disappears when the $T$ approaches and/or surpasses $\sim$ 200 K.
(\textbf{b}) $\rho_{\text{c}}$ as a function of $H$ from a third crystal at $T = 1.8$ K, and for several angles $\theta$ between $H$ and the \emph{c}-axis. Notice how the negative
magnetoresistivity observed at low fields is progressively suppressed as $\theta$ increases, becoming strongly positive. Nevertheless, the mechanism leading to the
negative magnetoresistivity is observed to overpower the orbital one at higher fields and higher angles. (\textbf{c}) Kohler plot for all the temperature dependent 
$\rho_{\text{c}}/\rho_0$. Red line is a fit of $\Delta \rho_{\text{c}}/ \rho_0$ to $H^{-1}$.}
\end{center}
\end{figure*}
\begin{figure}[b]
\begin{center}
\includegraphics[width = 6 cm]{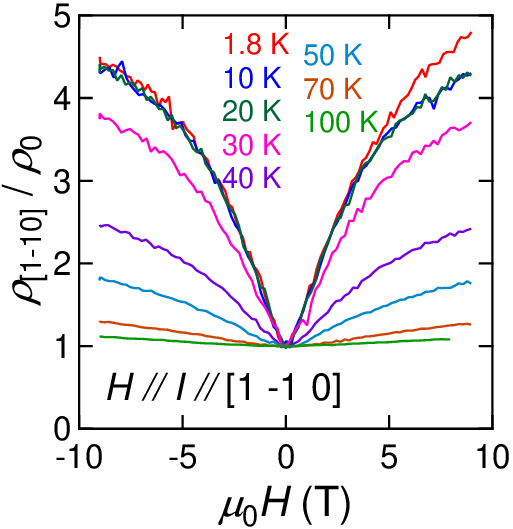}
\caption{\textbf{Longitudinal magnetoresistive for fields along the planes.} In-plane longitudinal resistivity $\rho_{[1\overline{1}0]}$ normalized by its zero field value $\rho_0$
as a function of the field applied along the $[1\overline{1}0]$ direction, for a PdCoO$_2$ single-crystal and for several temperatures.
Notice the absence of negative magnetoresistivity.}
\end{center}
\end{figure}
\begin{figure*}[htb]
\begin{center}
\includegraphics[width = 11 cm]{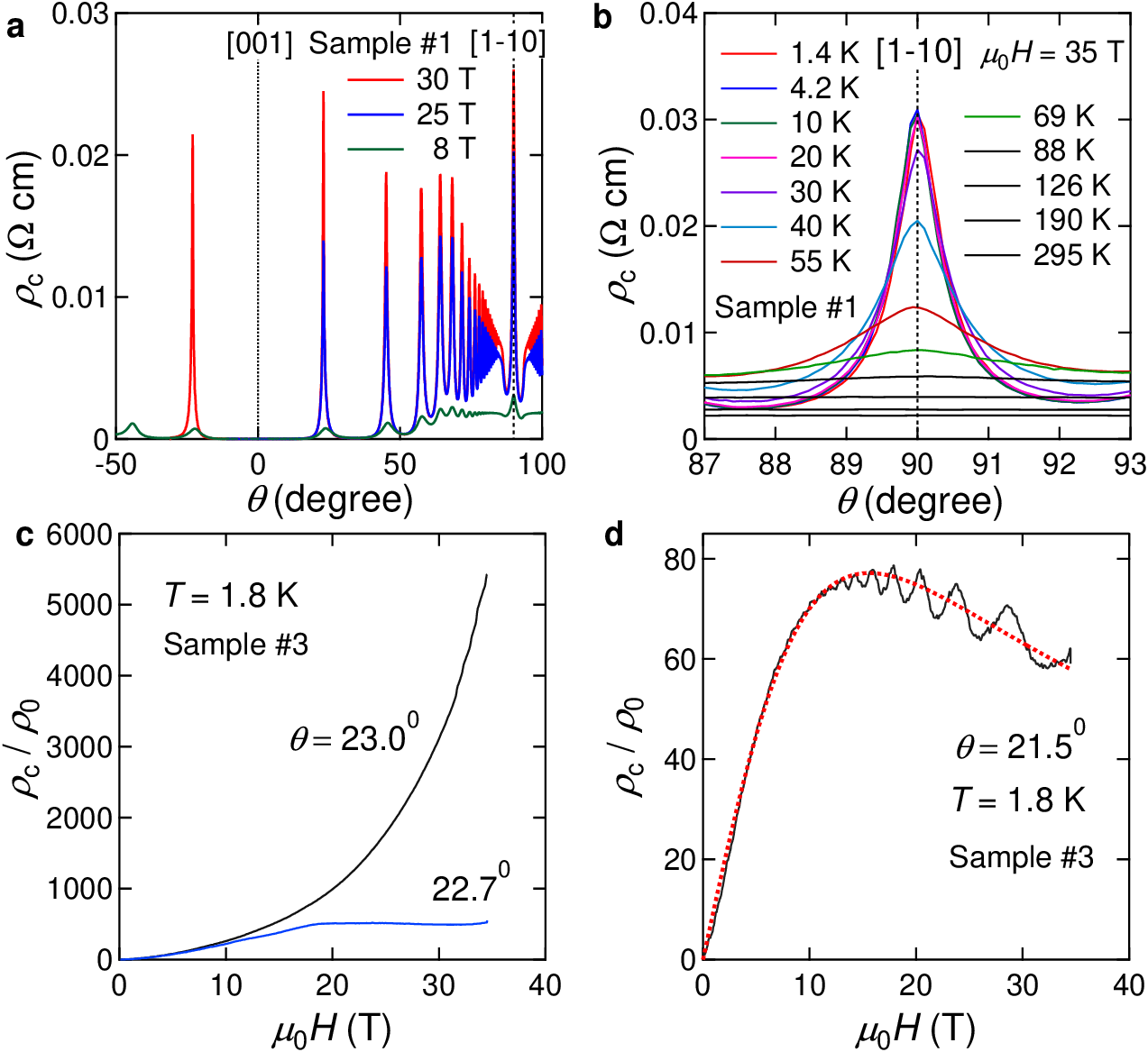}
\caption{\textbf{Angular magnetoresistance oscillations for a PdCoO$_2$ single-crystal}.
(\textbf{a}) inter-planar magnetoresistivity $\rho_{\text{c}}$ for a PdCoO$_2$ single-crystal as a function of the angle $\theta$ between the [001] inter-planar direction and the
external field $H$. The pronounced peaks observed as a function of $\theta$ are the so-called Yamaji-effect peaks\cite{yamaji}. (\textbf{b}) Inter-layer coherence peak observed
for fields nearly along the inter-planar direction, which indicates an extended Fermi surface along the inter-layer direction\cite{singleton}. From the width $\Delta \theta$ of
the peak at half maximum, one can estimate the value of the inter-layer transfer integral $t_{\text{c}} = 2.79$ meV from Eq. 1 in the main text. (\textbf{c}) Inter-planar resistivity
$\rho_{\text{c}}$ as a function of $H$ at $T = 1.8$ K and for two angles, i.e. the Yamaji value $\theta_{\text{{n=1}}} = 23.0^{\circ}$ and $\theta = 22.7^{\circ}$. Notice how the pronounced
positive magnetoresistivity observed at $\theta_{\text{{n=1}}}$ is strongly suppressed when $H$ is rotated by just $\sim 0.3^{\circ}$ leading to magnetoresistance saturation.
(\textbf{d}) $\rho_{\text{c}}$ as a function of $H$ under $T = 1.8$ K and for $\theta = 21.5^{\circ}$. Notice how $\rho_{\text{c}}$, after increasing by several orders of magnitude, displays negative
magnetoresistivity at higher fields thus indicating a clear competition between the orbital and another mechanism which suppresses the magnetoresistivity. Dotted red line corresponds to a fit of $\rho_{\text{c}} = 1/\sigma_{\text{c}}^{-1} = (\sigma_0 + \alpha H + \beta/H)^{-1}$.}
\end{center}
\end{figure*}

The evolution of the longitudinal magnetoresistance with temperature is depicted in Figure 3a. $\rho_{\text{c}}$ is seen to decrease by a factor surpassing 60 \% for fields approaching 9 T and for all temperatures below 30 K. Figure 3b displays $\rho_{\text{c}}(H)/ \rho_0$ as a function of the angle $\theta$ between $H$ and the \emph{c}-axis at a temperature $T = 1.8$ K, for a third single-crystal. For $\theta > 10^{\circ}$ the pronounced positive magnetoresistance observed at low fields, due to a classical orbital magnetoresistive effect, is overpowered at higher fields by the mechanism responsible for the negative
magnetoresistivity. This behavior is no longer observed within this field range when $\theta$ is increased beyond $\sim 20^{\circ}$.  Figure 3c shows a Kohler
plot, i.e. $\Delta \rho_{\text{c}}/\rho_0$ as a function of $H$ normalized by $\rho_0$. As seen in Fig. 3c, all curves collapse on a single curve indicating that
a particular transport mechanism dominates even at high temperatures where phonon scattering is expected to be strong.
The red line is a fit to $H^{-1}$, indicating that $\rho_{\text{c}}^{-1} = \sigma_{\text{c}} \propto H$ at lower fields.

Fig. 4 shows the longitudinal magnetoresistance $\rho_{[1\overline{1}0]}(H)/\rho_0$ for fields and currents along the $[1\overline{1}0]$-axis. For this orientation
the charge carriers follow open orbits along the axis of the cylindrical Fermi surface instead of quantized cyclotronic orbits. In contrast to $\Delta \rho_{\text{c}}/\rho_0$, but similarly to the longitudinal magnetoresistivity of ultra-clean elemental metals \cite{pippard,pippard2}, $\rho_{[1\overline{1}0]}(H)/\rho_0$ is
observed to increase and saturate as a function of $H$. This further confirms that conventional mechanisms, e.g. impurities, magnetism, etc.,
are not responsible for the negative longitudinal magnetoresistivity observed in $\Delta \rho_{\text{c}}/\rho_0$.

Figure 5a shows $\rho_{\text{c}}$ as a function of the angle $\theta$ between the field and the \emph{c}-axis, for three different field values: 8, 25 and 30 T. $\rho_{\text{c}} (\theta)$ displays the characteristic structure displayed by quasi-two-dimensional metals, namely a series of sharp peaks at specific angles
$\theta_{\text{n}} = \arctan(\pi(n-1/4)/ck_{\text{F}}^{\|})$ called Yamaji angles (where $n$ is an integer, $c$ is the inter-planar distance, and $k_{\text{F}}^{\|}$ is the projection
of the Fermi wave number on the conduction plane), for which all cyclotronic orbits on the FS have an identical orbital area~\cite{yamaji}.
In other words, the corrugation of the FS no longer leads to a distribution of cross-sectional areas, as if the corrugation has been effectively suppressed.
As discussed below, in terms of the energy spectrum, this means that the Landau levels become non-dispersive at the Yamaji angles~\cite{Goswami,kurihara}, hence one no longer has Fermi points. The sharp peak at $\theta=90^{\circ}$ is attributed to coherent electron transport along small closed orbits on the sides of a corrugated cylindrical
Fermi surface\cite{singleton,hanasaki}. The width of this peak $\Delta \theta$, shown in Fig. 5b for several temperatures, allows us to estimate the interlayer transfer
integral $t_{\text{c}}$\cite{uji},
\begin{equation}
\Delta \theta \approx \frac{2k_{\text{F}} t_{\text{z}} d}{E_{\text{F}}}\rightarrow t_{\text{c}} \approx \frac{\Delta\theta E_{\text{F}}}{2k_{\text{F}} d}
\end{equation}
assuming a simple sinusoidal Fermi surface corrugation along the $k_{\text{z}}$-direction. Here, the inter-planar separation is $d=c/3$, since there are three conducting Pd planes
per unit  cell, each providing one conducting hole and therefore leading to 3 carriers per
unit cell. This value is consistent with our Hall-effect measurements (not included here).  The full width at half maximum of the peak at $90^{\circ}$ is
$\Delta\theta \simeq 0.78^{\circ}$ and $E_{\text{F}}$ is given by $\hbar^2 k_{\text{F}}^2/2 \mu = 2.32$ eV, therefore one obtains $t_{\text{c}} = 2.79$ meV or $\simeq 32.4 \text{ K}$.
Figure 5c displays $\rho_{\text{c}}$ as a function of $H$ for two angles; the Yamaji angle $\theta_{\text{n=1}} = 23.0^{\circ}$ and $\theta = 22.7^{\circ}$, respectively.
As seen, $\rho_{\text{c}}(H)$ for fields along $\theta_{\text{n=1}}$ displays a very pronounced positive magnetoresistance, i.e. $\rho_{\text{c}}/\rho_0$ increases by
$\sim 550,000$ \% when $H$ is swept from 0 to 35 T. But at $H=35$ T $\rho_{\text{c}}/\rho_0$ decreases by one order of magnitude as $H$ is rotated by just $\sim 0.3^{\circ}$
with respect $\theta_{\text{n=1}}$. Furthermore, as seen in Fig. 5d, at higher fields $\rho_{\text{c}}$ displays a crossover from a very pronounced and positive to a
negative magnetoresistance, resulting from a small increment in $\theta$ relative to $\theta_{\text{n=1}}$. This is a very clear indication for two competing mechanisms,
with negative magnetoresistivity overcoming the orbital-effect when the orbitally-averaged inter-layer group velocity (or the transfer integral $t_{\text{c}}$) becomes finite
at $\theta \neq \theta_{\text{n}}$. We emphasize that for a conventional and very clean metal, composed of a single Fermi surface sheet, the magnetoresistivity should either
be $\propto H^2$ \cite{ross} or saturate as seen in quasi-two-dimensional metals close to the Yamaji angle \cite{yagi}, or in Figs. 2a and 2b for fields
below $\sim 15$ T. This is illustrated by the Supplementary Fig. 1 which contrasts our experimental observations with predictions based on semi-classical transport models which
correctly describe the magnetoresistance of layered organic metals in the vicinity of the Yamaji angle. In contrast, as illustrated by the dotted red line in Fig. 5d
$\rho_{\text{c}}(H)$ can be well-described by the expression $\rho_{\text{c}}(H)= \sigma_{\text{c}}^{-1}=(\sigma_0+\alpha H +\beta/H)^{-1}$. Here, the $\rho_{\text{c}} \propto H^{-1}$
term describes the negative magnetoresistivity as previously seen in Fig. 3, while the $\rho_{\text{c}} \propto H$ term describes the non-saturating linear magnetoresistance
predicted and observed for systems close to the quantum-limit \cite{abrikosov,silvertelluride,rosenbaum,rosenbaum2}. This expression describes $\rho_{\text{c}} (H, \theta)$
satisfactorily, except at the Yamaji angle where both terms vanish. In the neighborhood of $\theta_{n}$ the addition of a small $\rho_{\text{c}} \propto H^2$ term improves the fit,
with its pre-factor increasing as $\theta_{\text{n}}$ is approached. $\rho_{\text{c}}$ also displays Shubnikov de Haas oscillations at small (and strongly $\theta$-dependent)
frequencies which were not previously detected in Ref. \onlinecite{hicks}. As discussed in Ref. \onlinecite{kartsovnik} these slow oscillations, observed only in the interlayer
magnetoresistance of layered metals, originate from the warping of the FS. In the Supplementary Fig. 2, we show how these frequencies disappear when the group velocity vanishes at $\theta_{\text{n}}$.
\begin{figure}[htbp]
\begin{center}
\includegraphics[width = 7.8 cm]{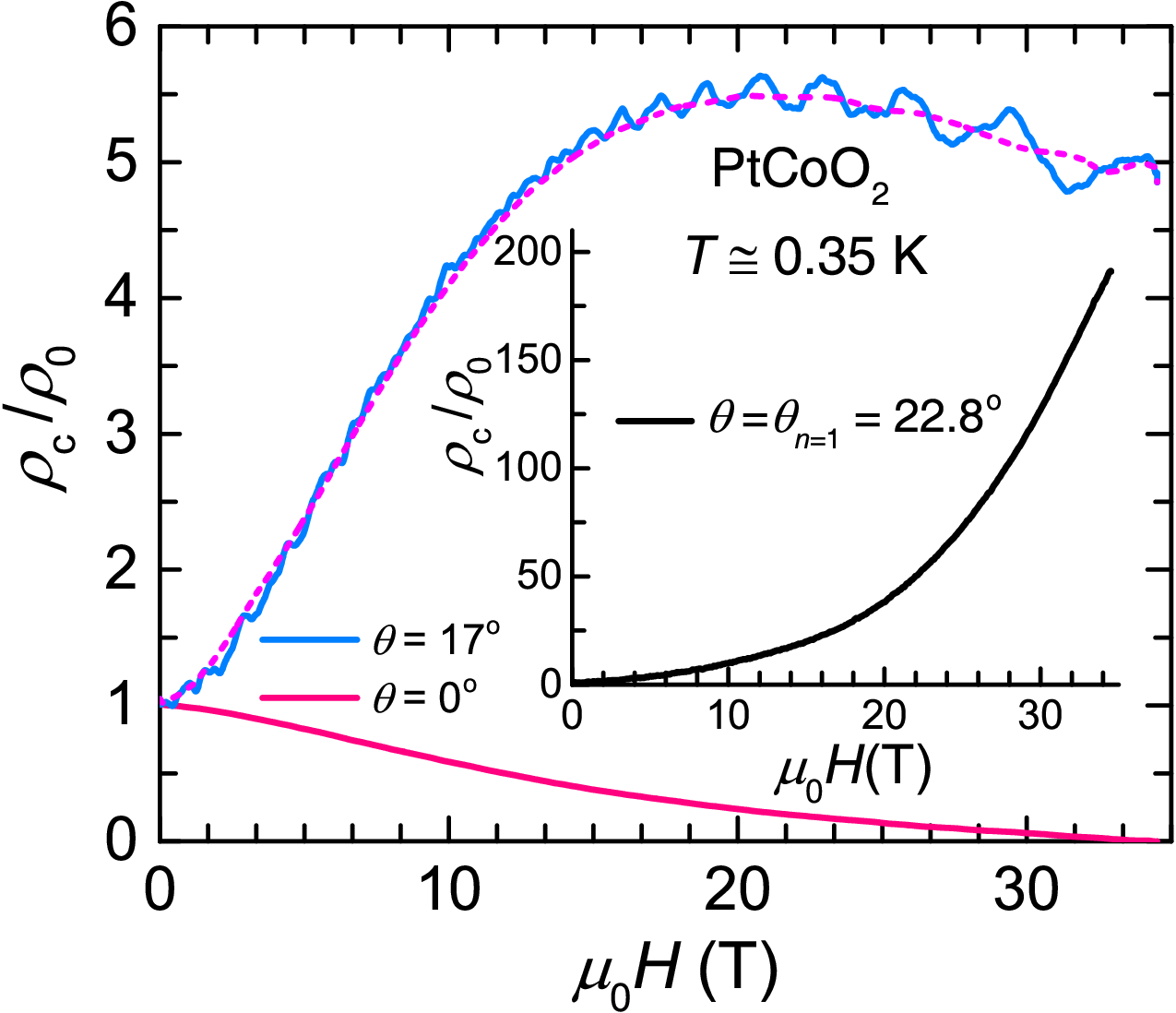}
\caption{\textbf{Negative longitudinal magnetoresistance in PtCoO$_2$}. Inter-planar resistivity $\rho_{\text{c}}$ normalized by its zero-field
value $\rho_0$ for a PtCoO$_2$ single-crystal at a temperature $T=0.35$ K, and as a function of the magnetic field $\mu_0 H$ applied along two angles with respect to
the \emph{c}-axis, respectively $\theta = 0^{\circ}$ (pink line) and $17^{\circ}$ (blue line). Dashed magenta line describes the smoothly varying background.
Inset: $\rho_{\text{c}}/\rho_0$ as a function of $\mu_0 H$ applied along the first Yamaji angle $\theta_{\text{n}} = 22.8^{\circ}$.}
\end{center}
\end{figure}

Significantly, this effect does not appear to be confined to PdCoO$_2$. Figure 6 presents an overall evaluation of the longitudinal magnetoresistance of isostructural PtCoO$_2$, while 
Supplementary Fig. 3 displays the observation of impurity dependent negative magnetoresistivity in the correlated perovskite Sr$_2$RuO$_4$.  As shown in Fig. 6, PtCoO$_2$ presents a 
pronounced negative longitudinal magnetoresistivity either for $\textbf{\text{j}} \parallel \textbf{\text{H}} \parallel$ \emph{c}-axis or for $H$ close to an Yamaji angle 
($\textbf{\text{j}}$ is the current density). It also presents a very pronounced and 
non-saturating magnetoresistivy for fields applied along the Yamaji angle. For both systems the magnetoresistivity does not follow a single power law as a function of $H$. In fact, 
as shown in the Supplementary Fig. 4, at $\theta_{\text{n}}$ the magnetoresistivity of the (Pt,Pd)CoO$_2$ system follows a $H^2$ dependence
for $ H \lesssim 15$ T. At intermediate fields $\rho(H)$ deviates from the quadratic dependence,
recovering it again at subsequently higher fields. Since Kohler's rule implies that $\Delta \rho/ \rho_0 \propto (\mu_0H/\rho_0)^2$, we argue that the observed increase in slope 
would imply a field-dependent reduction in scattering by impurities (see Supplementary Fig. 4 and Supplementary note 4). The precise origin of this suppression in scattering remains 
to be identified. Nevertheless, the enormous and positive magnetoresistivity observed for fields along $\theta_{\text{n}}$ seems consistent with a simple scenario, i.e., a extremely 
clean system(s) whose impurity scattering weakens with increasing magnetic field. In Sr$_2$RuO$_4$, the negative longitudinal magnetoresistivity
is observed only in the cleanest samples and for angles within $10^{\circ}$ away from the \emph{c}-axis. This compound is characterized by three corrugated cylindrical
Fermi surface sheets, each leading to a distinct set of Yamaji-angles, making it impossible to completely suppress the inter-planar coupling at specific Yamaji angle(s).

\section{Discussion}
Negative magnetoresistivity is a common feature of ferromagnetic metals near their Curie temperature, or of samples having dimensions comparable to their electronic mean free path
where the winding of the electronic orbits under a magnetic field reduces the scattering from the surface. It can also result from the field-induced suppression of weak localization or from the field-induced suppression of spin-scattering/quantum-fluctuations as seen in \emph{f}-electron compounds\cite{bin}. None of the compounds described in this article are near a magnetic instability, nor do they contain significant amounts of magnetic impurities or disorder to make them prone to weak localization. The magnitude of this anomalous magnetoresistivity, coupled to its peculiar angular dependence, are in fact enough evidence against any of these conventional mechanisms. Below, we discuss an alternative scenario based on the so-called axial anomaly which in our opinion explains most of our observations.

The axial anomaly is a fundamental concept of relativistic quantum field theory, which describes the violation of separate number conservation laws of left and right handed massless
chiral fermions in odd spatial dimensions due to quantum mechanical effects~\cite{peskin, fujikawa}. In particular, when three dimensional massless Dirac or Weyl fermions are placed
under parallel electric and magnetic fields, the number difference between the left and the right handed fermions is expected to vary with time according to the Adler-Bell-Jackiw
formula~\cite{adler, bell} \begin{equation}\partial_t \left(n_{\text{R}}-n_{\text{L}}\right)=\frac{e^2EB}{2\pi^2 \hbar^2} \label{eq1}.\end{equation} Here, $n_{\text{R/L}}$ are the number operators for the
right and the left handed Weyl fermions, with the electric and the magnetic field strengths respectively given by $E$ and $B$. The Dirac fermion describes the linear touching of
two-fold Kramers degenerate conduction and valence bands at isolated momentum points in the Brillouin zone. The axial anomaly was initially proposed
to produce a large, negative longitudinal magnetoresistance, for a class of gapless semiconductors, for which the low energy band structure is described by
massless Weyl fermions \cite{nielsen}. The reason for the negative magnetoresistance is relatively straightforward. Assume that in very clean system the impurity
scattering potential can be described by a smoothly decaying function $\Omega (\textbf{Q})$ which peaks at $\textbf{Q}=0$ (e.g. Gaussian) hence producing scattering at low angles.
Once an external $H$ is applied, the system would require backscattering at the larger momentum transfer $\textbf{Q}_\text{W}$ which connects both Weyl-points in \emph{k}-space
in order to equilibrate the field-induced net number imbalance among right and left handed fermions produced by the anomaly. Given the functional form of $\Omega (\textbf{Q})$,
backscattering at $\textbf{Q}_\text{W}$ will be considerably weaker than backscattering at $\textbf{Q}=0$. Therefore, the role of the anomaly is to suppress
backscattering causing a net decrease in the resistivity under an applied magnetic field. Recent theoretical proposals for Weyl semi-metals~\cite{vishwanath, xu,burkov,weng} followed by experimental confirmation \cite{lv,yang} have revived the interest in the experimental confirmation of the axial anomaly through efforts in detecting negative longitudinal magnetoresistivity~\cite{aji,son,kim1,vishwanathsid,burkov1,kim2,huang}.
There are examples of three dimensional Dirac semi-metals ~\cite{ZX, Zahid, Cava}, which may be converted, through Zeeman splitting,
into a Weyl semi-metal. Examples include Bi$_{\text{1-x}}$Sb$_{\text{x}}$ at the band inversion transition point between topologically trivial and nontrivial insulators \cite{kim1},
and Cd$_3$As$_2$~\cite{ong}.
\begin{figure}[htbp]
\begin{center}
\includegraphics[width = 6.9 cm]{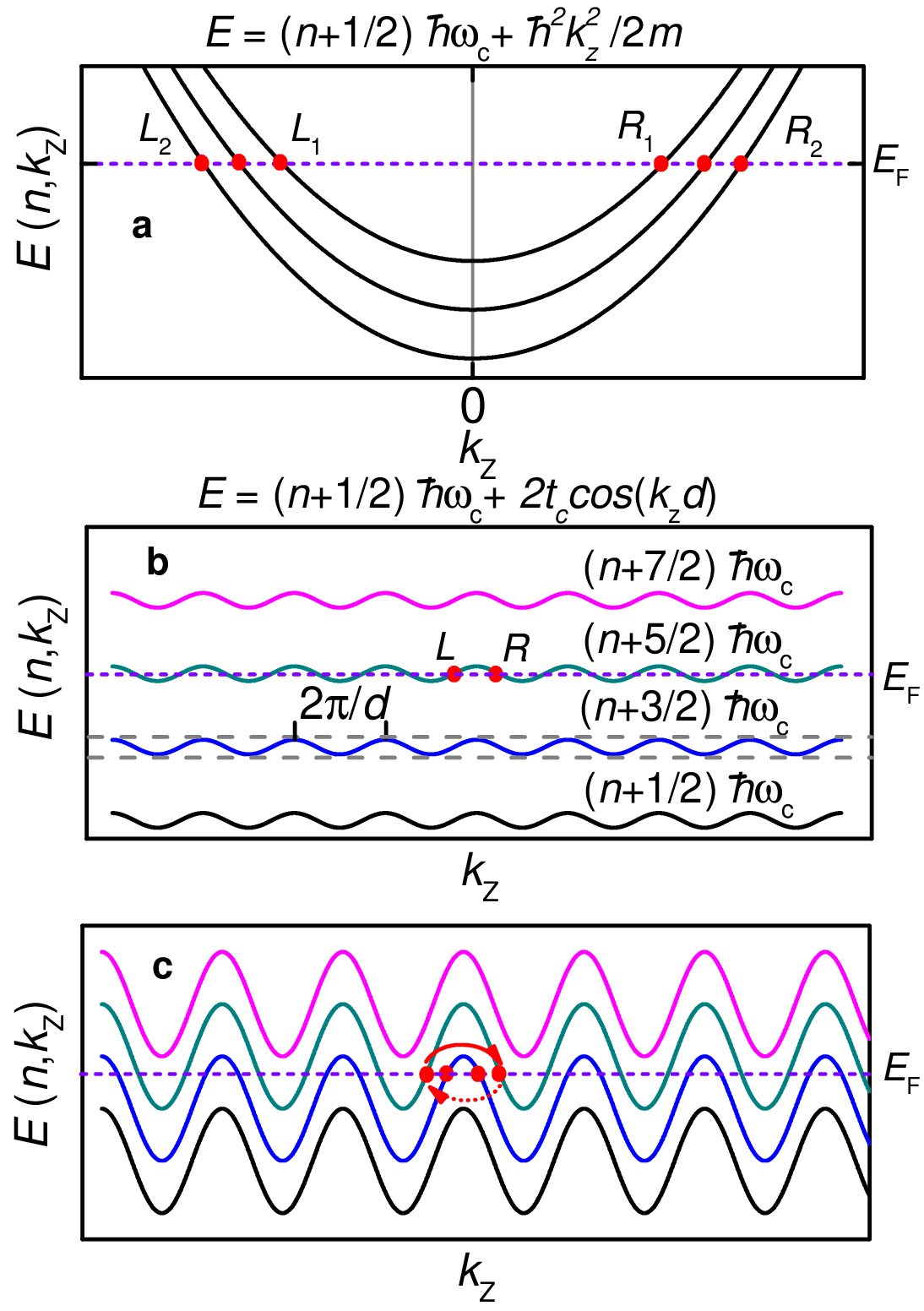}
\caption{\textbf{Field-induced electronic dispersion for metals of different dimensionality}.
(\textbf{a}) The dispersion of Landau levels for a conventional three-dimensional metal placed in an external magnetic field applied along the \emph{z}-direction.
Due to the underlying parabolic dispersion, each Landau level disperses quadratically as a function of $k_\text{z}$, the momentum component along the applied field.
Each partially occupied Landau level intersects the Fermi energy $E_{\text{F}}$ at two Fermi points, as indicated by the red dots. In the vicinity of the two Fermi points
located at $k_{\text{z}}=\pm k_{\text{F,n}}$ for the \emph{n}-th partially filled Landau level, the quasiparticles disperse linearly with opposite group velocities
$v_{\pm,\text{n}}=\pm \hbar k_{\text{F,n}}/\mu$ where $\mu$ is the effective mass. The $\pm$ signs of the group velocity respectively define the chirality of the right and the left
moving one dimensional fermions. (\textbf{b}) In contrast, for quasi-two-dimensional metals the Landau levels possess a periodic dispersion relation as a function of
$k_{\text{z}}$ due to the tight binding term $2t_{\text{c}}\cos (k_{\text{z}} d)$, with interlayer hopping strength and spacing respectively given by $t_{\text{c}}$ and $d$.
Within the first Brillouin zone defined as $-\pi/d<k_{\text{z}}<\pi/d$, each partially filled Landau level again gives rise to a pair of one dimensional fermions of opposite
chirality around the Fermi points. The situation depicted here corresponds to $4t_{\text{c}} < \hbar \omega_{\text{c}}$, or when only one Landau level is partially filled.
(\textbf{c}) Landau levels for $4t_{\text{c}} > \hbar \omega_{\text{c}}$ or when multiple Landau levels are partially occupied and each of them gives rise to a pair
of chiral fermions.}
\end{center}
\end{figure}

In analogy with the predictions for the axial anomaly between Weyl points, here we suggest that our observations might be consistent with the emergence of the axial anomaly among the Fermi points of a field-induced, one-dimensional electronic dispersion \cite{Goswami}. In effect, in the presence of a strong magnetic field, the quantization of cyclotron motion leads to discrete Landau levels with one dimensional dispersion and a degeneracy factor $eB/h$, see Figs. 7a, b, and c. Consider the low energy description of a one dimensional electron gas, in terms of the right- and left-handed fermions obtained in the vicinity of the two Fermi points. In the presence of an external electric field $E$, the separate number conservation of these chiral fermions is violated  according to \begin{equation}\partial_t \left(n_{\text{R}}-n_{\text{L}}\right)=\frac{eE}{\pi \hbar},\label{eq2}\end{equation}where $n_{\text{R/L}}$ respectively correspond to the number operators of the right- and left-handed fermions~\cite{peskin,fujikawa}.  Each partially occupied Landau level leads to a set of Fermi points, and the axial anomaly for such a level can be obtained from Eq.~\ref{eq2}, after multiplying by $eB/h$. Therefore each level has an axial anomaly determined by Eq.~\ref{eq1}. When only one Landau level is partially filled, we have the remarkable universal result for the axial anomaly described by Adler-Bell-Jackiw formula of Eq.~\ref{eq1}. For a non-relativistic electron gas this would occur at the quantum limit. In contrast, this situation would naturally occur for Dirac/Weyl semi-metals, when the Fermi level lies at zero energy. Figure 6b describes the situation for a quasi-two dimensional electronic system upon approaching the quantum limit, or when the inter-planar coupling becomes considerably smaller than the inter Landau level separation (e.g. in the vicinity of the Yamaji angle). We emphasize that the observation of a pronounced, linear-in-field magnetoresistive component, as indicated by the fit in Fig. 5d, is a strong experimental evidence for the proximity of PdCoO$_2$ to the quantum limit upon approaching the Yamaji angle.  Therefore, we conclude that the axial anomaly should be present in every three dimensional conducting system, upon approaching the quantum limit. Explicit calculations indicate that the axial anomaly would only cause negative magnetoresistance for predominant forward scattering produced by ionic impurities~\cite{Argyres,Goswami}. $\rho(H) \propto H^{-1}$ as observed here (Figs 3 and 5), would result from Gaussian impurities \cite{Goswami}. As our experimental results show, PdCoO$_2$ is a metal of extremely high conductivity thus necessarily dominated by small angle scattering processes and therefore satisfying the forward scattering criterion. In this metal the Landau levels disperse periodically as shown in Figs. 7b and 7c, depending on the relative strength of the cyclotron energy
$\hbar\omega_{\text{c}}=\hbar eB/m$ with respect to the inter-layer transfer integral $t_{\text{c}}$. The condition $4t_{\text{c}}> \hbar \omega_{\text{c}}$ is satisfied when $H$
roughly exceeds 100 T. For this reason Fig. 7c, with multiple partially occupied Landau levels, describes PdCoO$_2$ for fields along the \emph{c}-axis or for arbitrary
angles away from the Yamaji ones. Nevertheless, one can suppress the Fermi points by aligning the field along an Yamaji angle, and this should suppress the associated axial anomaly.
As experimentally seen, the suppression of the Fermi points suppresses the negative magnetoresistivity indicating that the axial anomaly is responsible for it.

In summary, in very clean layered metals we have uncovered a very clear correlation between the existence of Fermi points in a one-dimensional dispersion and the observation of an anomalous negative magnetoresistivity. The suppression of these points leads to the disappearance of this effect. This indicates that the axial anomaly, and related negative magnetoresistivity, would not be contingent upon the existence of an underlying three dimensional Dirac/Weyl dispersion. Instead, our study in PdCoO$_2$, PtCoO$_2$ and Sr$_2$RuO$_4$, which are clean metals with no Dirac/Weyl dispersion at zero magnetic field, indicates that the axial anomaly and its effects could be a generic feature of metal(s) near the quantum limit. Nevertheless, the detection of negative magnetoresistivity would depend on the underlying scattering mechanisms, i.e. observable only in those compounds which are clean enough to be dominated by elastic forward scattering caused by ionic impurities~\cite{Argyres,Goswami}. In a generic metal with a high carrier density, it is currently impossible to reach the quantum limit; for the available field strength many Landau levels would be populated, thus producing a myriad of Fermi points. In this regard, extremely pure layered metals such as (Pd,Pt)CoO$_2$ are unique, since by just tilting the magnetic-field in the vicinity of the Yamaji angle, one can achieve the condition of a single, partially filled Landau level as it would happen at the quantum limit. Therefore, the observation in all three compounds, of a pronounced negative magnetoresistivity for fields and currents precisely aligned along the \emph{c}-axis is somewhat puzzling. This situation necessarily leads to the population of a multitude of Landau levels and, although it does not necessarily preclude a role for the axial anomaly, it does involve scattering among a number of Fermi points at distinct Landau levels. Nevertheless, the suppression of this negative magnetoresistivity for
fields precisely aligned along the Yamaji angles indicates unambiguously, that the electronic structure at the Fermi level is at the basis for its underlying mechanism. The situation
is somewhat analogous to that of the Weyl semi-metals which are characterized by a number of Weyl-points in the First Brillouin zone \cite{weng}, and apparently with all Weyl-points contributing to
its negative longitudinal magnetoresistivity \cite{huang}. Hence, our results suggest that the axial anomaly among pairs of chiral Fermi points may play a role in ultra-clean systems even when they are located far
from the quantum limit.

Finally, notice that negative longitudinal magnetoresistivity is also seen in kish graphite at high fields, which is characterized by ellipsoidal electron- and hole-like Fermi surfaces, upon approaching the quantum limit and prior to the onset of a many body instability towards a field-induced insulating density-wave ground state \cite{behnia}. As discussed in Ref. \onlinecite{Goswami} the axial anomaly upon approaching the quantum limit may also play a role for the negative magnetoresistivities observed in ZrTe$_5$ \cite{li} and in $\alpha-$(ET)$_2$I$_3$ \cite{tajima} indicating that this concept, which is the basis of our work, is likely to be relevant to a number of physical systems, in particular  semi-metals.

\section{Methods}

\noindent
{\textbf{Crystal Synthesis.}} Single crystals of PdCoO$_2$ were grown through the following solid state reaction
PdCl$_2$ + 2CoO $\rightarrow$ PdCoO$_2$ + CoCl$_2$ with starting powders
of PdCl$_2$ (99.999 \%) and CoO (99.99+ \%). These powders were ground for
for up to 60 min and placed in a quartz tube. The tube was sealed in vacuum
and heated up to 930$^{\circ}$C in a horizontal furnace within 2 h, and subsequently
up to 1000$^{\circ}$C within 6 h, and then cooled down quickly to 580$^{\circ}$C in 1 or 2 h.
The tube is heated up again to 700$^{\circ}$C within 2 h, kept at 700$^{\circ}$C for 40 h, and
then cooled down to room temperature at 40$^{\circ}$C/h. Single crystals were
extracted by dissolving out CoCl$_2$ with hot ethanol.
Single-crystals sizes of approximately $2.8 \times 1.3 \times 0.3$ mm$^3$ were obtained.

\noindent
{\textbf{Single-crystal Characterization.} These were characterized by powder X-ray diffraction (XRD), energy dispersive
X-ray analysis (EDX), and electron probe microanalysis (EPMA).
The XRD pattern indicated no impurity phases. In the crystals measured for this study EPMA indicated
that the ratio of Pd to Co is 0.98:1 and that the amount of Cl impurities is $< 200$ ppm.

\noindent
\textbf{Experimental set-up.} Transport measurements were performed by using conventional four-terminal techniques in conjunction
with a Physical Properties Measurement System, a 18 T superconducting solenoid and a 35 T resistive-magnet, coupled to cryogenic facilities such as
$^3$He systems and variable temperature inserts.

\section{Acknowledgements}
We thank S.\ Das Sarma, V.\ Yakovenko, L.\ Balents, E.\ Abrahams and J.\ Pixley for useful discussions.
The NHMFL is supported by NSF through NSF-DMR-1157490 and the
State of Florida.  L.~B. is supported by DOE-BES through award DE-SC0002613.

\section{Author contributions}
NK performed the measurements and analyzed the data.
AK, ESC, DG, RB, JSB, SU, KS, TT, PMCR and NEH contributed to the collection of experimental data at high magnetic fields.
LB provided scientific guidance and PG the theoretical interpretation.  HT, SY and YM synthesized and characterized the single crystals. YI and MN performed electron probe microanalysis of the measured single-crystals to confirm their high degree of purity.
PG, NH and LB wrote the manuscript with the input of all co-authors.

\section{Additional information}

Supplementary information is available in the online version of the paper. Reprints and
permissions information is available online at www.nature.com/reprints.
Correspondence and requests for materials should be addressed to N. K. or L.B.

\section{Competing financial interests}
The authors declare no competing financial interests.


\begin{thebibliography}{0}
\bibitem{pippard} Pippard, A.\ B. \emph{Magnetoresistance in Metals}. Cambridge Studies in Low temperature Physics 2, Cambridge University Press (1989).\
\bibitem{pippard2} Pippard, A.\ B.\ Longitudinal Magnetoresistance. \emph{Proc. Roy. Soc.} \textbf{A282}, 464-484 (1964).
\bibitem{rosenbaum} Lee, M., Rosenbaum, T.\ F., Saboungi, M.\-L. \& Schnyders, H. S. Band-Gap Tuning and Linear Magnetoresistance in the Silver Chalcogenides, \emph{Phys. Rev. Lett.} \textbf{88}, 066602  (2002).\
\bibitem{abrikosov} Abrikosov, A. A., Quantum linear magnetoresistance, \emph{Europhys. Lett.} \textbf{49}, 789–793 (2000).\
\bibitem{silvertelluride} Zhang, W., \emph{et al}., Topological Aspect and Quantum Magnetoresistance of $\beta$-Ag$_2$Te. \emph{Phys. Rev. Lett.} \textbf{106}, 156808 (2011).\
\bibitem{ong} Liang, T. \emph{et al}. Ultrahigh mobility and giant magnetoresistance in the Dirac semi-metal Cd$_3$As$_2$. \emph{Nat. Mater.} \textbf{14}, 280-284 (2015).\
\bibitem{ali} Ali, M.\ N. \emph{et al}. Large, non-saturating magnetoresistance in WTe$_2$, Nature \textbf{514}, 205-208 (2014).\
\bibitem{weyl_semimetals} Huang,  S.-M. \emph{et al}. Theoretical Discovery/Prediction: Weyl Semimetal states in the TaAs material (TaAs, NbAs, NbP, TaP) class. \emph{Nat. Commun}. \textbf{6}, 7373 (2015).\
\bibitem{adler} Adler, S. Axial-Vector Vertex in Spinor Electrodynamics. \emph{Phys. Rev.} \textbf{177}, 2426-2438 (1969).\
\bibitem{bell} Bell, J. S. \& Jackiw, R. A PCAC puzzle: $\pi_0 \rightarrow \gamma \gamma$ in the $\sigma$-model. \emph{Nuovo Cimento A} \textbf{60}, 47-61 (1969).\
\bibitem{nielsen} Nielsen H. B. \& Ninomiya, M. The Adler-Bell-Jackiw Anomaly and Weyl Fermions in a Crystal \emph{Phys. Lett.} \textbf{130B}, 389-396 (1983).\
\bibitem{takatsu1} Takatsu, H.\ \emph{et al}. Roles of High-Frequency Optical Phonons in the Physical Properties of the Conductive Delafossite PdCoO$_2$. \emph{J. Phys. Soc. Jpn.} \textbf{76}, 104701 (2007).\
\bibitem{band1} Eyert, V., Fr\'{e}sard, R. \& Maignan, A. On the Metallic Conductivity of the Delafossites PdCoO$_2$ and PtCoO$_2$. \emph{Chem.\ Mater.} \textbf{20}, 2370-2373 (2008).\
\bibitem{band2} Seshadri, R., Felser, C., Thieme, K. \& Tremel, W. Metal-metal Bonding and Metallic Behavior in Some ABO$_2$ Delafossites. \emph{Chem.\ Mater.} \textbf{10}, 2189-2196 (1998).\
\bibitem{band3} Kim, K., Choi, H. C. \& Min, B. I. Fermi surface and surface electronic structure of delafossite PdCoO$_2$. Phys.\ Rev.\ B \textbf{80}, 035116 (2009).\
\bibitem{hicks} Hicks, C.\ W. \emph{et al}. Quantum Oscillations and High Carrier Mobility in the Delafossite PdCoO$_2$. \emph{Phys. Rev. Lett.} \textbf{109}, 11640 (2012).\
\bibitem{arpes} Noh, H.\-J.\ \emph{ et al}. Anisotropic Electric Conductivity of Delafossite PdCoO$_2$ Studied by Angle-Resolved Photoemission Spectroscopy. \emph{Phys.\ Rev.\ Lett.} \textbf{102}, 256404 (2009).\
\bibitem{takatsu2} Takatsu, H. \emph{et al}. Extremely Large Magnetoresistance in the Nonmagnetic Metal PdCoO$_2$. \emph{Phys. Rev. Lett.} \textbf{111}, 056601 (2013).\
\bibitem{Goswami} Goswami, P., Pixley, J. \& Das Sarma, S. Axial anomaly and longitudinal magnetoresistance of a generic three dimensional metal. \emph{Phys. Rev. B} \textbf{92}, 075205 (2015).\
\bibitem{antilocalization} Hikami, S., Larkin, A. I., \& Nagaoka, Y. Spin-Orbit Interaction and Magnetoresistance in the Two-Dimensional Random System. \emph{Prog. Theor. Phys.} \textbf{63},  707-710  (1980).\
\bibitem{Bergmann} Bergmann, G. Weak localization in thin films a time-of-flight experiment with conduction electrons. \emph{Phys. Rep.} \textbf{107}, 1-58 (1984).
\bibitem{ross} Moses, P.\ \& Mackenzie R.\ H. Comparison of coherent and weakly incoherent transport models for the interlayer magnetoresistance of layered Fermi liquids.\ \emph{Phys. Rev. B} \textbf{60}, 7998 (1999).\
\bibitem{yamaji} Yamaji, K.\ On the Angle Dependence of the Magnetoresistance in Quasi-Two-Dimensional Organic Superconductors. \emph{J.\ Phys.\ Soc.\ Jpn} \textbf{58}, 1520-1523 (1989).
\bibitem{kurihara} Kurihara, Y. A Microscopic Calculation of the Angular-Dependent Oscillatory Magnetoresistance in Quasi-Two-Dimensional Systems.
 \emph{J.\ Phys.\ Soc.\ Jpn} \textbf{61}, 975-982 (1992).
\bibitem{singleton} Singleton, J., \emph{et al}. Test for interlayer coherence in a quasi-two-dimensional superconductor. \emph{Phys. Rev. Lett.} \textbf{88}, 037001 (2002).\
\bibitem{hanasaki} Hanasaki, H., Kagoshima, S., Hasegawa T., Osada, T., \& Miura, N. Contribution of small closed orbits to magnetoresistance in quasi-two-dimensional conductors. \emph{Phys. Rev. B} \textbf{57}, 1336-1339 (1998).\
\bibitem{uji} Uji, S.\ \emph{et al}. Fermi surface and angular-dependent magnetoresistance in the organic conductor (BEDT-TTF)$_2$Br(DIA). \emph{Phys. Rev. B} \textbf{68}, 064420 (2003).\
\bibitem{yagi} Yagi, R., Iye Y., Osada, T., \& Kagoshima, S. Semiclassical Interpretation of the Angular-Dependent Oscillatory Magnetoresistance in Quasi-Two-Dimensional Systems.
\emph{J. Phys. Soc. Jpn.} \textbf{59}, 3069-3072 (1990).
\bibitem{rosenbaum2} Hu, J. \& Rosenbaum, T. F. Classical and quantum routes to linear magnetoresistance. \emph{Nat. Mater.} \textbf{7}, 697-700 (2008).
\bibitem{kartsovnik} Kartsovnik, M. V., Grigoriev, P. D., Biberacher, W., Kushch, N. D. \& Wyder, P. Slow Oscillations of Magnetoresistance in Quasi-Two-Dimensional Metals. \emph{Phys. Rev. Lett.} \textbf{89}, 126802 (2002).
\bibitem{bin} Zeng, B. \emph{et al}. CeCu$_2$Ge$_2$: Challenging our understanding of quantum criticality. \emph{Phys. Rev. B} \textbf{90}, 155101 (2014).\
\bibitem{peskin} Peskin, M. E. \& Schroeder, D. V., {\it An Introduction to Quantum Field Theory}, (Addison-Wesley, 1995).\
\bibitem{fujikawa} Fujikawa, K. \& Suzuki, H., {\it Path Integrals and Quantum Anomalies}, (Clarendon Press, 2004).\
\bibitem{vishwanath} Wan, X., Turner, A., Vishwanath, A. \& Savrasov, S. Y. Topological semimetal and Fermi-arc surface states in the electronic structure of pyrochlore iridates. \emph{Phys. Rev. B} \textbf{83}, 205101 (2011).\
\bibitem{xu} Xu, G., Weng, H., Wang, Z., Dai, X. \& Fang, Z. Chern Semimetal and the Quantized Anomalous Hall Effect in HgCr$_2$Se$_4$. \emph{Phys. Rev. Lett.} \textbf{107}, 186806 (2011).\
\bibitem{burkov} Burkov A. A. \& Balents, L. Weyl Semimetal in a Topological Insulator Multilayer. \emph{Phys. Rev. Lett.} \textbf{107}, 127205
(2011).\
\bibitem{weng} Weng, H. M. \emph{et al}. Weyl Semimetal Phase in Noncentrosymmetric Transition-Metal Monophosphides. \emph{Phys. Rev. X} \textbf{5}, 011029 (2015).
\bibitem{lv} Lv, B. Q. \emph{et al}. Observation of Weyl nodes in TaAs. \emph{Nat. Phys.} \textbf{11}, 724-727 (2015).
\bibitem{yang} Yang, L. X. \emph{et al}. Weyl semimetal phase in the non-centrosymmetric compound TaAs. \emph{Nat. Phys.} \textbf{11}, 728-732 (2015).
\bibitem{aji}Aji, V., Adler-Bell-Jackiw anomaly in Weyl semi-metals: Application to Pyrochlore Iridates. \emph{Phys. Rev. B} \textbf{85}, 241101 (2012).\
\bibitem{son} Son, D. T. \& Spivak, B. Z. Chiral anomaly and classical negative magnetoresistance of Weyl metals. \emph{Phys. Rev. B} \textbf{88}, 104412 (2013).\
\bibitem{kim1} Kim, H.-J. \emph{ et al}. Dirac versus Weyl Fermions in Topological Insulators: Adler-Bell-Jackiw Anomaly in Transport Phenomena. \emph{Phys. Rev. Lett.} \textbf{111}, 246603 (2013).\
\bibitem{vishwanathsid}Parameswaran, S. A., Grover, T., Abanin, D. A., Pesin, D. A. \&  Vishwanath, A. Probing the Chiral Anomaly with Nonlocal Transport in Three-Dimensional Topological Semimetals. \emph{Phys. Rev. X}, \textbf{4}, 031035 (2014).\
\bibitem{burkov1} Burkov, A. A. Chiral Anomaly and Diffusive Magnetotransport in Weyl Metals. \emph{Phys. Rev. Lett.} \textbf{113}, 247203 (2014)
\bibitem{kim2} Kim, K.-S., Kim, H.-J. \& Sasaki, M. Boltzmann equation approach to anomalous transport in a Weyl metal. \emph{Phys. Rev. B} \textbf{89}, 195137 (2014).\
\bibitem{huang} Huang, X. C. \emph{et al}. Observation of the Chiral-Anomaly-Induced Negative Magnetoresistance in 3D Weyl Semimetal TaAs. \emph{Phys. Rev. X} \textbf{5}, 031023 (2015).
\bibitem{ZX}Liu, Z. K. \emph{ et al}. Discovery of a Three-Dimensional Topological Dirac Semimetal Na$_3$Bi, \emph{Science} \textbf{343}, 864-867 (2014).
\bibitem{Zahid}Neupane, M. \emph{ et al}. Observation of a three-dimensional topological Dirac semimetal phase in high-mobility Cd$_3$As$_2$. \emph{Nat. Commun.} \textbf{5}, 3786 (2014).\
\bibitem{Cava} Borisenko, S. \emph{ et al}. Experimental Realization of a Three-Dimensional Dirac Semimetal. \emph{Phys. Rev. Lett.} \textbf{113}, 027603 (2014).
\bibitem{Argyres}Argyres, P. N. \& Adams, E. N. Longitudinal Magnetoresistance in the Quantum Limit. \emph{Phys. Rev.} \textbf{104}, 900-908 (1956).\
\bibitem{behnia} Fauqu\'{e}, B. \emph{et al.} Two Phase Transitions Induced by a Magnetic Field in Graphite. \emph{Phys. Rev. Lett.} \textbf{110}, 266601 (2013).
\bibitem{li} Li, Q. \emph{et al}. Observation of the chiral magnetic effect in ZrTe$_5$. Preprint at http://arxiv.org/abs/1412.6543 (2014).
\bibitem{tajima} Tajima, N., Sugawara, S., Kato, R., Nishio, Y. \&  Kajita, K. Effects of the Zero-Mode Landau Level on Inter-Layer
Magnetoresistance in Multilayer Massless Dirac Fermion Systems. \emph{Phys. Rev. Lett.} \textbf{102}, 176403 (2009).
\end{thebibliography}
\end{document}